\documentclass[nofootinbib,prd,preprintnumbers,superscriptaddress,aps]{revtex4}
\usepackage{amsmath}
\usepackage{amsfonts}
\usepackage{graphicx}
\usepackage{amssymb}

\usepackage[colorlinks=true,linkcolor=blue,citecolor=teal,urlcolor=blue]{hyperref}
\usepackage{xcolor}
\numberwithin{equation}{section}

\usepackage{physics}
\usepackage{bbold}
\usepackage{cases}
\usepackage{braket}
\usepackage[T1]{fontenc}
\usepackage{mathrsfs}
\usepackage{enumerate}
\usepackage{bm}
\usepackage{multirow}
\usepackage{comment}
\usepackage{mathrsfs}

\makeatother

\begin{document}
\title{Constraining Modified Mass-to-Horizon Cosmology Through Primordial Inflationary Observables}


\author{A. Sheykhi}
\email{asheykhi@shirazu.ac.ir} \affiliation{Department of Physics,
College of Science, Shiraz University, Shiraz 71454, Iran}
\affiliation{Biruni Observatory, College of Science, Shiraz
University, Shiraz 71454, Iran}

\author{G.~G.~Luciano}
\email{giuseppegaetano.luciano@udl.cat} \affiliation{Applied
Physics Section of Environmental Science Department,  Escola
Polit\`ecnica Superior, Universitat de Lleida, Av. Jaume II, 69,
25001 Lleida, Spain}

\author{A. Benkrane}
\email{abdelhakim.benkrane@univ-ouargla.dz}
\affiliation{University Kasdi Merbah Ouargla, Laboratoire LRPPS,
Ouargla 30000, Algeria}

\date{\today}

\begin{abstract}
We investigate slow-roll inflation in a modified cosmological
framework inspired by a generalized mass-to-horizon relation
(MHR), $M=\gamma {c^2 L^n}/{G}$, where $n$ is a real parameter and
$\gamma$ a dimensional constant. Using Padmanabhan's emergence
paradigm, we derive the modified Friedmann equations for a flat
FRW universe and analyze the dynamics of a canonical scalar field
(inflaton) under the slow-roll approximation. We study the
resulting inflationary phenomenology for power-law and Starobinsky
potentials. For power-law potentials, the MHR modification fails
to reconcile these models with current CMB constraints on $r$ and
$n_s$. In contrast, Starobinsky inflation exhibits significant
sensitivity to deviations from $n=1$. A perturbative analysis
($n=1+\Delta$) yields corrections to inflationary observables. We
observe that the scalar power-spectrum normalization, under a
fixed-Starobinsky prescription, imposes the stringent constraint
$0.960 \lesssim n \lesssim 1.040$ for $N=60$ efolds. This is
considerably tighter than spectral-index bounds. Our results
establish inflation, particularly Starobinsky-like models, as a
sensitive probe of generalized horizon thermodynamics and
departures from standard MHR scaling.\\

Keywords: Mass-to-Horizon Cosmology; Inflation; Starobinsky
potentials.
\end{abstract}

\maketitle

\section{Introduction\label{Intro}}
The inflationary paradigm provides a compelling framework for
describing the dynamics of the early Universe, offering elegant
solutions to several fundamental cosmological problems, including
the horizon, flatness, and monopole problems, while also providing
a natural mechanism for generating the primordial perturbations
that seed the observed large-scale structure. In its conventional
formulation, inflation is driven by a scalar field, commonly
referred to as the inflaton, whose potential energy dominates the
cosmic energy density and induces a phase of accelerated
expansion. Despite its remarkable success, the standard
inflationary framework leaves several important questions open,
particularly concerning the underlying physics of the inflaton,
its connection to the subsequent cosmic evolution, and the nature
and origin of dark energy. These open issues have motivated the
exploration of alternative and extended inflationary scenarios
that can provide a more comprehensive description of the early and
late-time Universe while remaining consistent with current
cosmological observations.

Recent advancements in cosmology have prompted the exploration of
alternative frameworks that integrate thermodynamic principles
with gravitational dynamics. One such approach is the modified MHR
cosmology, which posits a relationship between the mass of the
universe and its cosmological horizon. This framework not only
provides a novel perspective on the nature of gravity but also
offers insights into the thermodynamic properties of spacetime.
The MHR serves as a bridge between the microscopic structure of
spacetime and its macroscopic cosmological behavior, suggesting
that the universe's evolution is deeply intertwined with its
thermodynamic characteristics \cite{Goh1,Goh2}.

The idea that gravity may emerge from the thermodynamic properties
of spacetime has attracted considerable attention over the past
two decades. In his seminal work, Jacobson \cite{Jac} sharpened
this connection by demonstrating that the Einstein field equations
can be interpreted as an equation of state. In particular, by
associating an entropy proportional to the area with each local
Rindler horizon, together with the corresponding Unruh temperature
experienced by a uniformly accelerated observer \cite{Unruh1976},
and imposing the Clausius relation $\delta Q=T\delta S$, one can
recover the Einstein field equations with a cosmological constant
\cite{Jac}. This result relies on the area law for horizon entropy
established in the context of black-hole thermodynamics
\cite{Bekenstein1973,Hawking1975}, and suggests that the Einstein
equations can be viewed as macroscopic consistency conditions
arising from a local thermodynamic description of spacetime. This
thermodynamic perspective was subsequently extended to cosmology,
where applying the first law of thermodynamics to the apparent
horizon of a Friedmann-Robertson-Walker (FRW) universe leads to
the standard Friedmann equations~\cite{Cai1,Cai2,Shey1,Shey2}.

A natural extension of this paradigm is to consider modifications
of the horizon entropy. Since the entropy is proportional to the
horizon area only in the Bekenstein-Hawking framework, corrections
motivated by quantum gravity, non-extensive statistics, or
higher-curvature theories alter the Friedmann equations. For
instance, entropy corrections from Barrow \cite{JohnD}, Tsallis
\cite{Tsallis1,Tsallis2}, Kaniadakis \cite{Kan1,Kan2} and more
general frameworks
\cite{Odintsov2023,Luciano2026a,ourabah2024other} have been
studied in the context of inflation and dark energy
\cite{SheT,John,Lamb,Ava2,Luciano,Luciv,Emm2,
SheB2,Lym,Her,Dre,Luciano:2025elo,Luciano:2025hjn,Jizba:2024klq}.
Interestingly, extensions of the entropy-geometry relation may, in
some gravitational theories, require a non-equilibrium
thermodynamic description, in which the Clausius relation is
supplemented by an internal entropy-production term
\cite{Eling2006}.

Starting from the above premises, in this work we focus on the
implications of the modified MHR cosmology for slow-roll
inflation. We begin by deriving the modified Friedmann equations
using the emergence scenario, which connects the dynamics of
cosmic expansion to the degrees of freedom associated with the
universe's horizon. This derivation highlights the role of entropy
in shaping the cosmological evolution and provides a framework for
understanding how modifications to the mass-horizon relation can
influence inflationary dynamics. We then explore the conditions
for slow-roll inflation within this modified framework, examining
how the parameters of the MHR entropy affect the inflationary
potential and the behavior of the inflaton field. Our analysis
aims to elucidate the interplay between thermodynamic principles
and inflationary dynamics, offering new insights into the early
universe's evolution and the nature of dark energy.

Building on the result from \cite{Goh1} that the entropic force
depends critically on the form of the MHR, we take the logical
step of generalizing the MHR itself. We therefore propose and
investigate the following generalized relation \cite{Goh2}
\begin{equation}\label{MHR}
M=\gamma \frac{c^2}{G}L^n,
\end{equation}
where $n$ is a non-negative real number, and $\gamma>0$ is a
constant with dimensions $[L]^{1-n}$.  In order to recover the
standard formalism, we require
$\gamma\to1$ in the limit $n\to1$.

By combining the generalized mass-horizon relation (\ref{MHR})
with the Hawking temperature in the Clausius relation, we derive a
new entropy associated with the cosmological horizon as
\cite{Goh2}
\begin{equation}\label{Sh}
S_h=\gamma \frac{2n}{n+1}L^{n-1} S_{BH},
\end{equation}
where $S_{BH}$ is the usual Bekenstein-Hawking entropy which obeys
the area law \cite{Bekenstein1973,Hawking1975} and $L$ is the
cosmological radius. Recently, by employing Iyer-Wald's approach
\cite{Wald:1993nt,Iyer:1994ys}, the entropy \eqref{Sh} has been
reconstructed from a generalized $f(R)$ Lagrangian
\cite{Mondal:2026mqb}, following the formalism developed in
\cite{DAgostino:2024sgm}.

The paper is organized as follows. In Section \ref{Emer}, we
review the emergence of cosmic space based on the MHR entropy,
derive the modified Friedmann equations, and discuss their
cosmological consequences. In Section \ref{SRI}, we investigate
power-law inflation within the modified MHR background, deriving
the slow-roll dynamics and the corresponding observable
predictions, and confronting them with observational constraints.
Section \ref{Starob} is devoted to Starobinsky inflation in the
MHR framework, where we analyze the modified slow-roll dynamics
and the corresponding inflationary predictions. The final section
presents our closing remarks. Throughout the paper, we use units
$\hbar=k_B=c=1$ and define the reduced Planck mass as $M_{\rm
Pl}=1/\sqrt{8\pi G}$.
\section{Emergence of the cosmic space through MHR entropy \label{Emer}}
We consider a spatially homogeneous and isotropic
Friedmann--Robertson--Walker (FRW) universe, described by the line
element
\begin{equation}
ds^2=h_{\mu\nu}dx^\mu dx^\nu+\tilde{r}^2(d\theta^2+\sin^2\theta d\phi^2),
\end{equation}
where $a(t)$ denotes the scale factor, $\tilde{r}=a(t)r$ is the
areal radius, and $k$ is the spatial curvature parameter, with
$k=-1$, $0$, and $1$ corresponding to open, flat, and closed
spatial geometries, respectively. Here, we take $x^0=t$, $x^1=r$,
and $h_{\mu\nu}=\mathrm{diag}(-1,a^2/(1-kr^2))$. Among the various
cosmological horizons, the apparent horizon is particularly
relevant from a thermodynamic perspective and provides a natural
setting for studying gravitational thermodynamics in an FRW
spacetime \cite{Cai1,Cai2,Hay1,Gibbons:1977mu}. The radius of the
apparent horizon is determined by the condition
$h^{\mu\nu}\partial_{\mu}\tilde{r}_A\partial_{\nu}\tilde{r}_A=0$,
which implies that the gradient of the areal radius is null on the
apparent horizon. For an FRW universe, this condition yields
\cite{Hay1,Hay2,Bak}
\begin{equation}\label{radius}
\tilde{r}_A=\frac{1}{\sqrt{H^2+k/a^2}},
\end{equation}
where $H=\dot{a}/a$ is the Hubble parameter. We also propose the
entropy of the apparent horizon is in the form of the generalized
MHR entropy \cite{Goh1,Goh2},
\begin{equation}\label{Ent1}
S_{h}=\frac{2\pi n \gamma}{G(n+1)}\tilde {r}_{A}^{n+1},
\end{equation}
To derive the modified Friedmann equations from the generalized
MHR entropy in Eq.~(\ref{Ent1}), we follow the emergence scenario
proposed by Padmanabhan \cite{PadEm} (see also
\cite{Padmanabhan:2002sha} for an alternative description).
According to Padmanabhan, in a pure de Sitter universe
characterized by the Hubble constant $H$, the holographic
principle can be expressed as $N_{\rm sur}=N_{\rm bulk}$, where
$N_{\rm sur}$ and $N_{\rm bulk}$ denote the degrees of freedom on
the boundary and in the bulk, respectively. For our Universe,
which is asymptotically de Sitter, as supported by numerous
astronomical observations, Padmanabhan proposed that the cosmic
volume increases during an infinitesimal cosmic time interval $dt$
according to \cite{PadEm}
\begin{equation}
\frac{dV}{dt}\propto
\left(N_{\mathrm{sur}}-N_{\mathrm{bulk}}\right).
\label{dV1}
\end{equation}
For a spatially flat universe, Padmanabhan took the temperature
and volume to be $T=H/(2\pi)$ and $V=4\pi/(3H^3)$, respectively.
This choice follows from the fact that, in this case, our Universe
may be regarded as an asymptotically de Sitter spacetime.
Mathematically, Padmanabhan proposed \cite{PadEm}
\begin{equation} \label{dV}
\frac{dV}{dt}=G(N_{\mathrm{sur}}-N_{\mathrm{bulk}}).
\end{equation}
Following Padmanabhan, the notion was also extended to a nonflat
universe where it was shown that the Friedmann equations in
Einstein, Gauss-Bonnet and more general Lovelock gravity with any
spatial curvature can be derived by applying the emergence
scenario to the apparent horizon \cite{Sheyem}. It was argued that
in this case one should replace the Hubble radius ($H^{-1}$) with
the apparent horizon radius $ \tilde
{r}_{A}=1/{\sqrt{H^2+k/a^2}}$, which is a generalization of the Hubble
radius for $k\neq0$. The generalization of Eq. (\ref{dV}) for a
nonflat universe was proposed as \cite{Sheyem}
\begin{equation}\label{dV1bis}
\frac{dV}{dt}=G\frac{\tilde {r}_{A}}{H^{-1}}
\left(N_{\mathrm{sur}}-N_{\mathrm{bulk}}\right).
\end{equation}
The temperature associated with the apparent horizon is assumed to
be \cite{CaiEm}
\begin{equation}\label{T2}
T=\frac{1}{2\pi  \tilde {r}_{A}}.
 \end{equation}
We note that, within an infinitesimal time interval $dt$, the
condition $\dot {R}\ll 2H \tilde {r}_{A}$ holds. This implies that
the radius of the apparent horizon remains effectively constant
during this brief period, akin to the conditions found in a de
Sitter universe \cite{CaiEm}. Padmanabhan's proposal indeed
connects the change in volume $dV$ during this infinitesimal
interval $dt$ of cosmic time to the degrees of freedom present.
This simplification leads to the well-known expression for the
horizon temperature (\ref{T2}). This assumption is crucial for
deriving the correct form of the Friedmann equations within
Padmanabhan's framework. Additionally, in the context of
Padmanabhan's emergent gravity paradigm, the relation for volume
change assumes that the system is in a state of near thermal
equilibrium at each infinitesimal time step. In this framework,
treating the horizon radius as effectively constant during this
short interval is both physically meaningful and aligns with the
principles of horizon thermodynamics in slowly varying spacetimes.

Using the entropy expression (\ref{Ent1}), the number of degrees
of freedom on the surface is given by
\begin{eqnarray} \label{Nsur2}
N_{\mathrm{sur}}=\frac{8\pi n \gamma}{G(3-n)} \tilde
{r}_{A}^{n+1}.
\end{eqnarray}
With this definition, the surface degrees of freedom are still
proportional to the generalized entropy $S_h$, but the
proportionality constant is chosen so that the resulting effective
gravitational constant matches the one derived from the first law
of thermodynamics. For $n=\gamma=1$, Eq. (\ref{Nsur2}) reduces to
the standard relation $N_{\rm sur}=4S_{h}$.

We also modify the Padmanabhan's proposal as
\begin{equation}\label{dV2}
\frac{d\tilde{V}_n}{dt}=G\frac{\tilde {r}_{A}}{H^{-1}}
\left(N_{\mathrm{sur}}-N_{\mathrm{bulk}}\right).
\end{equation}
where the effective volume is defined as $\tilde {V}_n=\alpha
\tilde {r}_{A} ^{n+2} $. Here $\alpha$ is a constant which for
latter convenience we choose it as
\begin{equation}\label{alpha}
\alpha=\frac{4\pi n \gamma }{n+2}.
\end{equation}
Clearly for $n=\gamma=1$, we have $\alpha=4\pi/3$ and $\tilde{V}_n
\rightarrow V=4\pi \tilde {r}_{A}^3/3$. The motivation for
choosing the effective volume $\tilde {V}_n$ instead of the usual
volume, comes from the fact that for the generalized MHR entropy
$S_h\sim \tilde {A}_{n}\sim \tilde {r}_{A}^{n+1} $. Thus, the
generalized volume corresponding to the generalized area $\tilde
{A}_{n}$ is expected to be $\tilde {V}_n \sim \tilde {r}_{A}
^{n+2}$.

We take the total energy contained within the apparent horizon as
the Komar energy,
\begin{equation}
E_{\mathrm{Komar}}=|(\rho +3p)|V,  \label{Komar}
\end{equation}
The number of degrees of freedom of the matter field in the bulk
is determined using the equipartition law of energy ($k_B=1$),
\begin{equation}
N_{\mathrm{bulk}}=\frac{2|E_{\mathrm{Komar}}|}{T}.  \label{Nbulk1}
\end{equation}
Combining this relation with Eq. (\ref{Komar}) and assuming, in an
expanding universe, $\rho+3p<0$, we find
\begin{equation}
N_{\rm bulk}=-\frac{16 \pi^2}{3}  \tilde {r}_{A}^4 (\rho+3p).
\label{Nbulk}
\end{equation}
Substituting relations (\ref{Nsur2}) and (\ref{Nbulk}) in
assumption (\ref{dV2}), after simplifying, we arrive at
\begin{eqnarray}
\frac{\alpha (n+2)}{4\pi H}\tilde {r}_{A}^{n-4}\dot{\tilde
{r}}_{A}-\frac{2n\gamma}{3-n}\tilde {r}_{A}^{n-3}=\frac{4\pi G
}{3}(\rho+3p). \label{Frgb11}
\end{eqnarray}
If we multiply both side of Eq. (\ref{Frgb11}) by factor
$2\dot{a}a$, after some algebra and using continuity equation,
$\dot{\rho}+3H(\rho+p)=0$, we reach
\begin{equation}\label{Frgbd2}
\left(\frac{2 n\gamma}{3-n}\right)\frac{d}{dt} \left(a^2 \tilde
{r}_{A}^{n-3} \right)=\frac{8 \pi G }{3} \frac{d}{dt}(\rho a^2).
\end{equation}
Integrating yields
\begin{equation}\label{Frgb3}
\left(H^2+\frac{k}{a^2}\right)^{(3-n)/2} = \frac{8\pi G_{\rm
eff}}{3} (\rho+\rho_{\Lambda}).
\end{equation}
where in the last step, we have used relation (\ref{radius}). Here
the effective gravitational constant is defined as
\begin{equation}\label{Geff}
G_{\rm eff}=\frac{(3-n)G}{2n\gamma},
\end{equation}
and $\rho_{\Lambda}=\Lambda/(8\pi G_{\rm eff})$, where $\Lambda$
serves as an integration constant that can be interpreted as the
cosmological constant. Note that $G_{\rm eff}>0$ and hence $n<3$.
When $n=\gamma=1$, we have $G_{\rm eff}\rightarrow G$ and the
Friedmann equation reduces to the standard case. For a flat
universe ($k=0$), the Friedmann equations can be written as
\begin{equation}\label{Fri}
H^{3-n} = \frac{8\pi G_{\rm eff}}{3} (\rho+\rho_{\Lambda}).
\end{equation}

If we define the density parameters as
\begin{equation}\label{Dens1}
\Omega_{m} =\frac{\rho_m}{\rho_{\rm cri}}, \  \  \  \
\Omega_{\Lambda} =\frac{\rho_{\Lambda}}{\rho_{\rm cri}},  \  \  \
\ \rho_{\rm cit}=\frac{3 H^{3-n}}{8\pi G_{\rm eff}},
\end{equation}
then the Friedmann equation can be rewritten as
\begin{equation}\label{Fri0}
\Omega_{m}+\Omega_{\Lambda}=1.
\end{equation}
Solving Eq. (\ref{Fri}), for the Hubble parameter, we find
\begin{eqnarray}\label{H}
&&H(z)=H_0\left[\Omega_{m,0}(1+z)^3+1-\Omega_{m,0}\right]^{\frac{1}{3-n}},
\end{eqnarray}
where $1+z=a^{-1}$ is the redshift, $\rho_{m}=\rho_{m,0} a^{-3}$,
where $\rho_{m,0}$ is the present matter density, and in the last
step, we have used relation (\ref{Fri0}). Let us note that the
evolution of $H$ depends only on the parameter $n$ and is
independent of $\gamma$. Cosmological model based on generalized
MHR entropy, implications for structure growth and primordial
gravitational waves have been explored in \cite{Luci1,ShMH}.

To sum up, we have derived the modified Friedmann equation
inspired by the generalized MHR entropy using the framework of
emergent gravity proposed in \cite{PadEm} and developed in
\cite{Sheyem}. In what follow we explore slow-roll inflation in
the context of modified cosmology inspired by MHR entropy.

\section{Power-law inflation in MHR cosmology}\label{SRI}
In this section, we investigate how the modified Friedmann
equations derived from the generalized MHR entropy affect the
dynamics of slow-roll inflation. We consider a single scalar field
$\phi$ (the inflaton) with potential $V(\phi)$ as the driver of
inflation.

For a flat universe ($k=0$) during inflation, we neglect the
cosmological constant term since inflation occurs at early times.
From Eq. (\ref{Fri}), the modified Friedmann equation becomes
\begin{equation}\label{Fr1}
H^{3-n} = \frac{8\pi G_{\text{\rm eff}}}{3} \rho_{\phi},
\end{equation}
The energy density and pressure for a homogeneous scalar field are
\begin{equation}\label{rhop}
\rho_{\phi} = \frac{1}{2}\dot{\phi}^2 + V(\phi), \quad p_{\phi} =
\frac{1}{2}\dot{\phi}^2 - V(\phi).
\end{equation}
The continuity equation for the inflaton field is
\begin{equation}\label{KG1}
\dot{\rho}_{\phi} + 3H(\rho_{\phi} + p_{\phi}) = 0,
\end{equation}
which leads to the standard Klein-Gordon equation
\begin{equation}\label{KG}
\ddot{\phi}+3H\dot{\phi}+V'(\phi) = 0,
\end{equation}
where $V'(\phi)=dV/d\phi$. Note that the Klein-Gordon equation
(\ref{KG1}) remains unchanged in form because it follows from
energy conservation, which is preserved in this modified
cosmology.

During slow-roll inflation, we assume the potential dominates over
the kinetic energy, $\dot{\phi}^2\ll V(\phi)$, and the field
accelerates slowly, $|\ddot{\phi}| \ll |3H\dot{\phi}|$,
$|\ddot{\phi}| \ll |V'(\phi)|$. Under these conditions, we have
$\rho_{\phi}\approx V(\phi)$, and the Friedmann equation (\ref{Fr1})
simplifies as
\begin{equation}\label{Fr2}
H \approx \left[\frac{8\pi G_{\text{\rm eff}}}{3}
V(\phi)\right]^{1/(3-n)},
\end{equation}
while the Klein-Gordon equation (\ref{KG}) reduces to
\begin{equation}
\label{KG2}
3H\dot{\phi} \approx -V'(\phi).
\end{equation}

Under the slow-roll approximation the inflaton dynamics are
governed by Eqs.~(\ref{Fr2}) and~(\ref{KG2}). To characterise the
deviation from de Sitter expansion we introduce the standard
slow-roll parameters
\begin{equation}\label{epet}
\varepsilon \equiv -\frac{\dot{H}}{H^{2}}, \qquad \qquad \eta
\equiv \frac{V''}{3H^2}.
\end{equation}
Using the modified Friedmann equation (\ref{Fr1}) together with
the continuity equation, we compute \(\dot{H}\). Differentiating
(\ref{Fr1}) and using \(\dot{\rho}_{\phi}=-3H\dot{\phi}^{2}\)
gives
\begin{equation}
(3-n)H^{2-n}\dot{H}=-8\pi G_{\mathrm{eff}} H\dot{\phi}^{2},
\end{equation}
which yields
\begin{equation}
\dot{H} = -\frac{8\pi
G_{\mathrm{eff}}}{3-n}\,H^{n-1}\dot{\phi}^{2}.
\end{equation}
Thus, the first slow-roll parameter becomes
\begin{equation}
\varepsilon = -\frac{\dot{H}}{H^{2}} = \frac{8\pi
G_{\mathrm{eff}}}{3-n}\,H^{n-3}\dot{\phi}^{2}.
\end{equation}
During slow-roll, we have \(\rho_{\phi}\approx V(\phi)\), and using
Eq. (\ref{Fr2}) yields
\begin{equation}\label{eps}
\varepsilon=\frac{3}{3-n}\,\frac{\dot{\phi}^{2}}{V}.
\end{equation}
The slow-roll condition \(\varepsilon\ll1\) is thus equivalent to
\(\dot{\phi}^{2}\ll\frac{3-n}{3}V\). Clearly, for \(n=1\), we
recover the standard relation \(\varepsilon =
\frac{3}{2}\dot{\phi}^{2}/V\).

Using the Klein-Gordon equation (\ref{KG2}) together with
Eq.(\ref{Fr2}), we obtain a potential-dependent form
\begin{equation}\label{eps1}
\varepsilon = \frac{1}{3(3-n)}\left(\frac{3}{8\pi
G_{\mathrm{eff}}}\right)^{\frac{2}{3-n}}
{V^{\prime}}^{2}\,V^{-\frac{5-n}{3-n}}.
\end{equation}
In the slow-roll approximation, the second slow-roll parameter is
defined as (\ref{epet}). A straightforward calculation using the
equations of motion leads to
\begin{equation}\label{eta1}
\eta = \frac{V''}{3} \left( \frac{3}{8\pi G_{\mathrm{eff}}}
\right)^{\frac{2}{3-n}} V^{-\frac{2}{3-n}}.
\end{equation}
The number of e-folds before the end of inflation is
\begin{equation}\label{N}
N = \int_{t_c}^{t_f} H\,dt = \int_{\phi_{c}}^{\phi_f}
\frac{H}{\dot{\phi}}\,d\phi.
\end{equation}
where $t_i$ and $t_f$ represent the beginning and the end of the
inflationary phase, respectively. Since the observable properties
of primordial perturbations are determined at the horizon
crossing, it is convenient to set the initial time of inflation to
this moment, i.e. $t_i = t_c$. In terms of the inflaton field,
this corresponds to $\phi_i = \phi_c$. Note that $\phi_c =
\phi(t_c)$ is the value of the inflaton field at horizon crossing,
and $\phi_f = \phi(t_f)$ denotes its value at the end of
inflation.  Using Eqs. (\ref{Fr2}) and~(\ref{KG2}), we arrive at
\begin{equation}
\label{N1}
N = \int_{\phi_f}^{\phi_c} 3\left(\frac{8\pi
G_{\mathrm{eff}}}{3}\right)^{\frac{2}{3-n}}
\frac{V^{\frac{2}{3-n}}}{V'}\,d\phi.
\end{equation}

\subsection{Power-law potential}

To illustrate the physical implications, we consider the power-law
potential
\begin{equation}\label{V1}
V(\phi)=V_{0}\,\phi^{p}, \qquad p>0.
\end{equation}
The parameter $p$ denotes the power-law index and determines the
shape of the monomial inflaton potential. Monomial potentials
represent a prototypical class of large-field inflationary models,
historically associated with the chaotic inflation scenario
\cite{Linde:1983gd}. In particular, $p=2$ corresponds to the
quadratic potential of a massive scalar field,  while $p=4$
corresponds to a quartic self-interacting potential. Besides these
canonical examples, linear and fractional-power potentials, such
as $p=1$ and $p<1$, can arise in axion-monodromy and related
large-field constructions \cite{McAllister:2008h}. More generally,
monomial potentials with different values of $p$ provide useful
benchmarks for testing inflationary dynamics against CMB
observations \cite{Planck:2018jri}. The normalization $V_0$ can be
fixed by the amplitude of the primordial scalar power spectrum,
while it drops out of the leading-order expressions for the
inflationary observables considered below.

During slow-roll, the Klein-Gordon equation gives
\begin{equation}\label{dp2bis}
\dot{\phi}\simeq-\frac{V^{\prime}(\phi)}{3H}, \qquad
V^{\prime}(\phi)=pV_{0}\phi^{p-1}.
\end{equation}
Using the modified Friedmann equation (\ref{Fr2}), the Hubble rate
becomes
\begin{equation}\label{Fr3}
H\simeq\Bigl(\frac{8\pi
G_{\mathrm{eff}}}{3}V_{0}\phi^{p}\Bigr)^{1/(3-n)}.
\end{equation}
Therefore
\begin{equation}\label{dp3}
\dot{\phi}\simeq -\frac{pV_{0}\phi^{p-1}}{3}\, \Bigl(\frac{8\pi
G_{\mathrm{eff}}}{3}V_{0}\phi^{p}\Bigr)^{-1/(3-n)}.
\end{equation}
Inserting this expression into Eq.~(\ref{eps}), after some
algebra, we find
\begin{equation} \label{eps4}
\varepsilon=\frac{p^{2}}{3(3-n)}\left( \frac{6n\gamma
M_{\mathrm{Pl}}^2}{3-n} \right)^{\frac{2}{3-n}}
V_0^{\frac{1-n}{3-n}} \phi^{\frac{1-n}{3-n}p - 2}.
\end{equation}
where we have used Eq. \eqref{Geff}
along with $G=1/(8\pi M_{\mathrm{Pl}}^2)$.

The second slow-roll parameter \(\eta\) can be obtained from its
definition in the slow-roll approximation. For the power-law
potential \(V^{\prime\prime}=p(p-1)V_{0}\phi^{p-2}\), it is easy
to show that
\begin{equation}
\eta\simeq\frac{V^{\prime\prime}}{3H^{2}}\simeq\frac{p(p-1)}{3H^{2}}V_{0}\phi^{p-2}.
\end{equation}
Substituting the expression for \(H^{2}\), one obtains \(\eta\) as
a function of \(\phi\). The result is
\begin{equation}\label{eta2}
\eta = \frac{p(p-1)}{3} \left( \frac{3}{8\pi G_{\mathrm{eff}}}
\right)^{\frac{2}{3-n}} V_0^{1-\frac{2}{3-n}}
\phi^{\frac{1-n}{3-n}p-2}.
\end{equation}
Using the definition (\ref{Geff}), the final explicit form is given by
\begin{equation}\label{eta3}
\eta = \frac{p(p-1)}{3} \left( \frac{6n\gamma
M_{\mathrm{Pl}}^2}{3-n} \right)^{\frac{2}{3-n}}
V_0^{\frac{1-n}{3-n}} \phi^{\frac{1-n}{3-n}p - 2}.
\end{equation}
For the standard case \(n=\gamma=1\), this reduces to $\eta =
p(p-1)M^2_{Pl}/\phi^2$.

\subsection{Inflationary dynamics and observational predictions}
We define the end of inflation (and the corresponding number of
e-folds $N$) as the moment when $\varepsilon(\phi_f)\approx1$. Using Eq. Eq. (\ref{eps4}), $\varepsilon = C \phi^{\alpha}$,
where
\begin{equation}
C = \frac{p^{2}}{3(3-n)} \left( \frac{6n\gamma
M_{\mathrm{Pl}}^{2}}{3-n} \right)^{\frac{2}{3-n}}
V_{0}^{\frac{1-n}{3-n}},
\end{equation}
and
\begin{equation}
\alpha = \frac{1-n}{3-n}\,p - 2,
\end{equation}
then we can find
\begin{equation}
\phi_f = \left(\frac{1}{C}\right)^{1/\alpha}, \qquad (\alpha\neq
0).
\end{equation}
Explicitly,
\begin{equation}\label{phif1}
\phi_f = \left[ \frac{3(3-n)}{p^{2}} \left( \frac{3-n}{6n\gamma
M_{\mathrm{Pl}}^{2}} \right)^{\frac{2}{3-n}}
V_{0}^{\frac{n-1}{3-n}} \right]^{1/\alpha}.
\end{equation}
Using Eq. (\ref{N1}), the number of e-folds can be written as
\begin{eqnarray}
N = \int_{\phi_f}^{\phi_c} \frac{K}{p} V_0^{\frac{n-1}{3-n}}
\phi^{\beta}\,d\phi,
\end{eqnarray}
where we have defined
\begin{eqnarray}
K = 3\left(\frac{3-n}{6n\gamma
M_{\mathrm{Pl}}^{2}}\right)^{\frac{2}{3-n}}, \   \  \  \beta =
\frac{p(n-1)}{3-n} + 1 .
\label{Ndef}
\end{eqnarray}
In the following, we focus on the inflationary branch characterized by
$\phi_c\gg\phi_f>0$ and $A\equiv\beta+1>0$, which allows for a
consistent large-$N$ approximation as discussed below.

Integrating Eq. \eqref{Ndef}, we obtain
\begin{equation}
N 
=\frac{K}{pA}
V_0^{\frac{n-1}{3-n}}
\left(
\phi_c^{A} - \phi_f^{A}
\right).
\end{equation}
Since the end-of-inflation contribution $\phi_f^A$ is subleading with
respect to $\phi_c^A$, we can write
\begin{eqnarray}
N \approx
\frac{K}{pA}
V_0^{\frac{n-1}{3-n}}
\phi_c^{A}.
\end{eqnarray}
Solving for \(\phi_c\), after some algebra, we obtain
\begin{eqnarray}\label{phic}
\phi_c = \left[ \frac{pA N}{3} \left( \frac{6n\gamma
M_{\mathrm{Pl}}^{2}}{3-n} \right)^{\frac{2}{3-n}}
V_0^{-\frac{n-1}{3-n}}\right]^ {\frac{1}{A}}.
\end{eqnarray}
It is a matter of calculation to show that for $n=\gamma=1$, the
above expression reduces to $\phi_c=\sqrt{2pN}M_{\mathrm{Pl}}$,
which matches the standard result. The exact expression, including the contribution from $\phi_f$,
would give a subleading correction in the large-$N$ regime.

Next, we calculate the tensor-to-scalar ratio $r$ and the scalar
spectral index $n_s$. For this purpose, we first obtain the
slow-roll parameters at horizon crossing, namely
$\varepsilon_c=\varepsilon(\phi_c)$ and $\eta_c=\eta(\phi_c)$, as
functions of the number of e-folds $N$ for a power-law potential.
Using the relation for $\phi_c$, we obtain
\begin{eqnarray}\label{phic2}
\varepsilon_c &=& \frac{p}{p(n-1) + 2(3-n)} \times \frac{1}{N},\\
\eta_c &=& \frac{(p-1)(3-n)}{p(n-1) + 2(3-n)} \times \frac{1}{N}.
\end{eqnarray}
These satisfy $\eta_c ={\varepsilon_c(p-1)(3-n)}/{p}$. For \(n=1\),
they reduce to the standard results, \(\varepsilon_c = p/(4N)\)
and \(\eta_c = (p-1)/(2N)\).

The tensor-to-scalar ratio \(r\) and the scalar spectral index
\(n_s\) are given by \cite{Baumann}
\begin{eqnarray}\label{rns1}
 r&=&16\varepsilon_c\,,\\[2mm]
 n_s&=& 1-6\varepsilon_c+2\eta_c\,.
    \label{ns}
\end{eqnarray}
It is worth noting that, in the present analysis, the standard
leading-order slow-roll relations are retained as a
phenomenological prescription. This provides a well-defined and
conservative benchmark for assessing the consequences of the MHR
modification at the background level. Indeed, the departure from
the standard inflationary dynamics is entirely encoded in the
modified Friedmann equation and is consequently propagated to the
inflationary observables through the modified slow-roll
parameters. A fully self-consistent determination of the scalar
and tensor spectra would require a dedicated cosmological
perturbation analysis of the underlying MHR framework, including
the derivation of the corresponding mode equations and their
normalization. Such an extension is beyond the scope of the
present analysis and will be addressed in the future work. It is
then straightforward to obtain
\begin{eqnarray}\label{r}
r&=&\frac{16p}{n(p-2)+(6-p)} \times \frac{1}{N},\\
n_s&=&1-\frac{2\bigl(n(p-1)+3\bigr)}{n(p-2)+(6-p)}\times\frac{1}{N}.\label{ns2}
\end{eqnarray}
We stress that these expressions have been obtained within the
slow-roll and large-$N$ regime. For the physical branch $n>0$ and
$\gamma>0$, positivity of the effective gravitational coupling
requires $0<n<3$. Moreover, in this range, the condition $A>0$
assumed above is equivalent to $n(p-2)+(6-p)>0$.

To assess the observational viability of the model, we now fix the
number of e-folds to the representative value $N=60$ and consider
the monomial potentials with $p=2/3,\,1,\,2,$ and $4$ (see Tab. \ref{tab:n_constraints_powerlaw}). For each
choice of $p$, Eqs.~(\ref{r}) and (\ref{ns2}) then provide direct
constraints on the MHR parameter $n$. As discussed above, we
work on the physical branch $0<n<3$ and impose the condition $A>0$
required by the large-$N$ approximation. For all the representative
values of $p$ considered here, the latter condition is automatically
satisfied throughout the interval $0<n<3$ and therefore does not
introduce any additional restriction on $n$.
\begin{table}[t]
\centering
\renewcommand{\arraystretch}{1.45}
\setlength{\tabcolsep}{12pt}
\begin{tabular}{c c c c}
\hline\hline
$p$
&
$r<0.036$
&
$n_s=0.9649\pm0.0042$
&
Combined allowed range
\\
\hline

$\dfrac{2}{3}$
&
$0<n<0.296$
&
$2.154<n<2.654$
&
$\varnothing$
\\[1mm]

$1$
&
$\varnothing$
&
$1.764<n<2.455$
&
$\varnothing$
\\

$2$
&
$\varnothing$
&
$0.708<n<1.716$
&
$\varnothing$
\\

$4$
&
$\varnothing$
&
$\varnothing$
&
$\varnothing$
\\
\hline\hline
\end{tabular}

\caption{Constraints on the MHR parameter $n$ for representative
monomial potentials at $N=60$. }
\label{tab:n_constraints_powerlaw}
\end{table}
We compare the theoretical predictions with the observational
upper bound $r<0.036$ at $95\%$ C.L.~\cite{BICEPKeck2021} and with
the Planck 2018 determination $n_s=0.9649\pm0.0042$ at $68\%$
C.L.~\cite{Planck2018VI}. The corresponding constraints on the MHR
parameter $n$ are first determined separately for each observable
and then combined by identifying their common allowed range. The
resulting intersection is therefore understood as a compatibility
region with the adopted observational bounds, rather than as a
joint statistical confidence interval.

In particular, we find that,
for $p=2/3$ the tensor and scalar constraints separately select
nonempty intervals of the MHR parameter $n$, but these intervals do
not overlap: the tensor bound requires $0<n<0.296$, whereas the
scalar spectral index favors $2.154<n<2.654$. For $p=1$ and $p=2$,
the tensor bound alone excludes the entire physical interval
$0<n<3$, despite the existence of ranges compatible with the
constraint on $n_s$. Finally, for $p=4$, no value of $n$ in the
physical branch is compatible with either of the adopted
observational constraints. Consequently, for $N=60$, none of the
representative monomial potentials considered here admits a common
range of $n$ simultaneously compatible with the observational
constraints on $r$ and $n_s$.

\begin{figure}[t]
\centering
\begin{minipage}{0.49\textwidth}
    \centering
    \includegraphics[width=\linewidth]{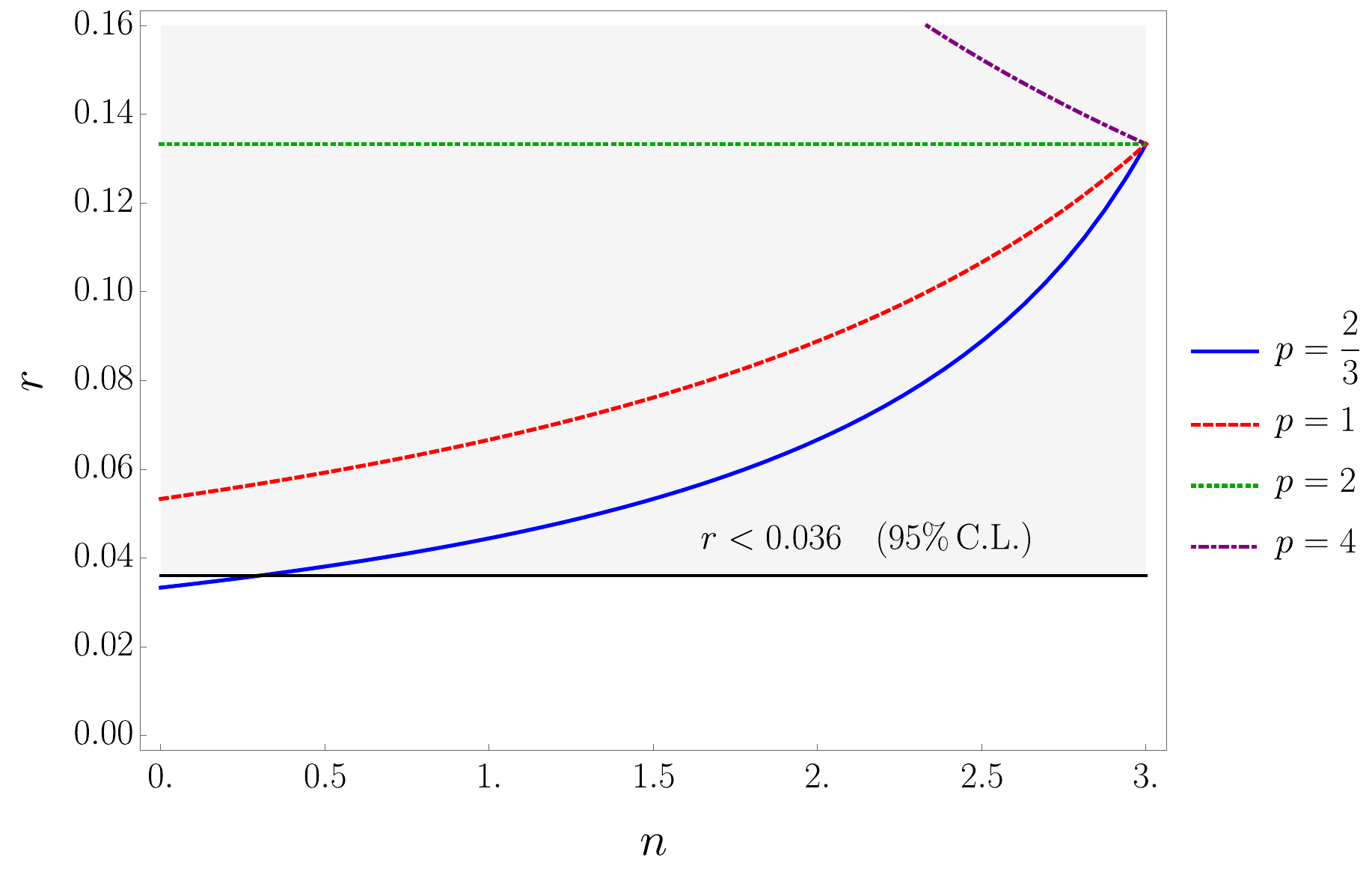}
\end{minipage}
\hfill
\begin{minipage}{0.49\textwidth}
    \centering
    \includegraphics[width=\linewidth]{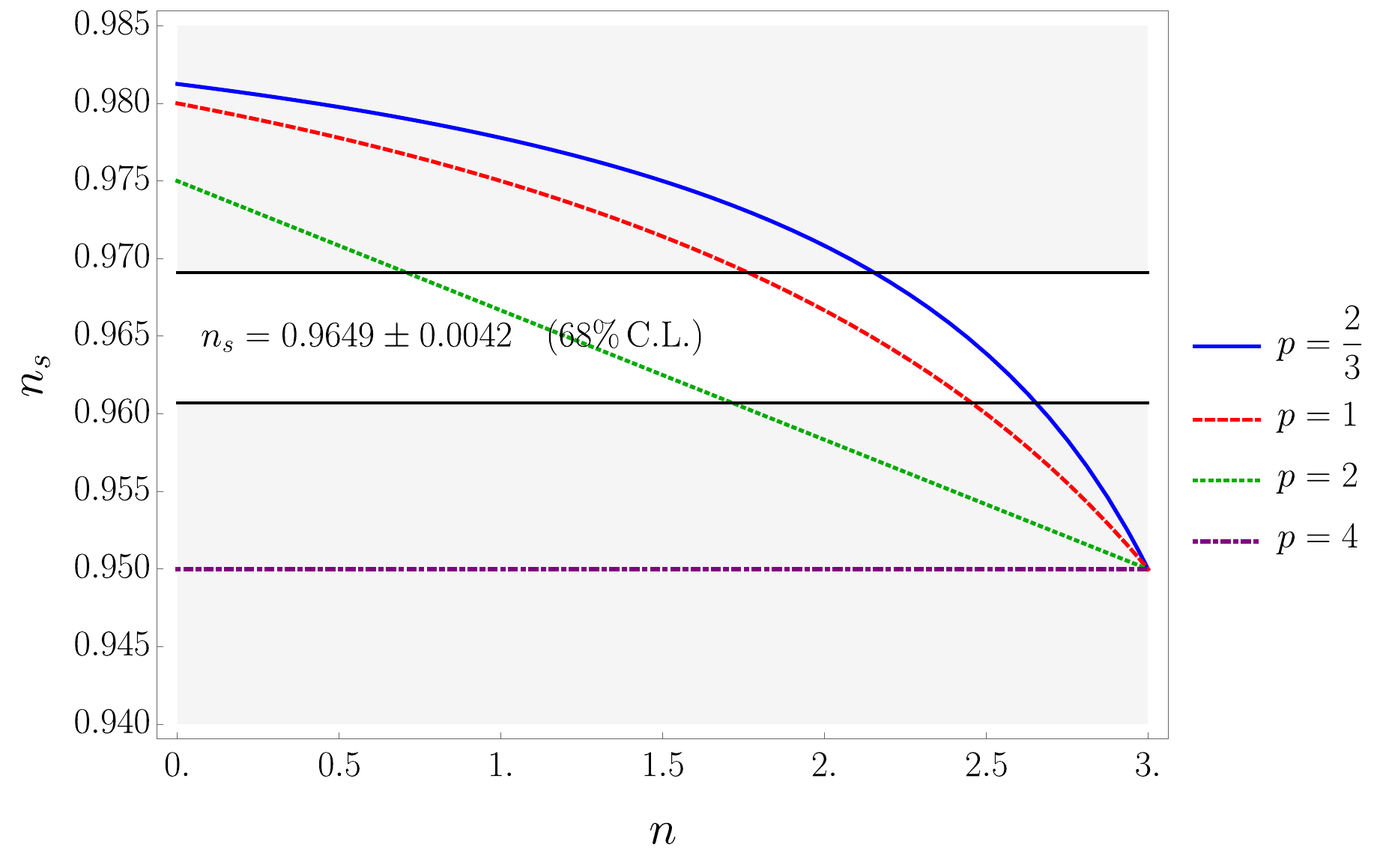}
\end{minipage}
\caption{Tensor-to-scalar ratio $r$ (left panel) and scalar spectral
index $n_s$ (right panel) as functions of the MHR parameter $n$ for
$N=60$ and the representative monomial potentials considered in the
text. The shaded regions denote the parameter space excluded by the
corresponding observational constraints (black solid lines).}
\label{Fig1}
\end{figure}

Figure~\ref{Fig1} shows the dependence of $r$ and $n_s$ on the MHR parameter
$n$ for the representative values of $p$. In both figures, the
shaded regions denote the parameter space excluded by the
corresponding observational constraint, whereas the unshaded regions
are observationally allowed.
These results therefore suggest that the modification of the
inflationary background dynamics induced by the MHR
horizon-entropy relation is not sufficient to reconcile power-law
inflation with the adopted CMB constraints. The observational
tension affecting these potentials in the standard scenario thus
appears to persist within the present MHR realization.
\section{Starobinsky inflationary model}
\label{Starob}
Let us now consider the Starobinsky inflationary potential~\cite{piattella2018lecture},
\begin{equation}
V(\phi)=\frac{3M^{2}M_{\mathrm{Pl}}^{2}}{4}
\left[
1-\exp\left(-\sqrt{\frac{2}{3}}\frac{\phi}{M_{\mathrm{Pl}}}\right)
\right]^{2},
\end{equation}
where $M_{\mathrm{Pl}}$ denotes the reduced Planck mass, which
sets the gravitational scale of the theory, while $M$
characterizes the inflationary energy scale. This potential
originates from the Starobinsky model, which can be viewed as an
effective gravitational theory resulting from the inclusion of
quadratic curvature corrections to the Einstein--Hilbert action.
Such corrections arise naturally from quantum effects associated
with matter fields propagating in a classical gravitational
background~\cite{piattella2018lecture}. Throughout this work, we
restrict our analysis to the large-field regime, $\phi \gg
M_{\mathrm{Pl}}$, where the slow-roll approximation is well
satisfied and inflation proceeds efficiently.

Using the modified Friedmann equation (\ref{Fr2}), the Hubble rate
becomes
\begin{equation}
\label{Fr3bis}
H=(2\pi)^{\frac{1}{3-n}}
\left[
G_{\mathrm{eff}}\,M^{2}M_{\mathrm{Pl}}^{2}
\left(
1-\exp\left[-\sqrt{\frac{2}{3}}\frac{\phi}{M_{\mathrm{Pl}}}\right]
\right)^{2}
\right]^{\frac{1}{3-n}}.
\end{equation}
From Eq. \eqref{KG2}, we thus obtain
\begin{equation}
\label{dp3bis2}
\dot{\phi}\simeq -\frac{
M^{2}M_{\mathrm{Pl}}\,
e^{-\sqrt{\frac23}\frac{\phi}{M_{\mathrm{Pl}}}}
\left(
1-e^{-\sqrt{\frac23}\frac{\phi}{M_{\mathrm{Pl}}}
}
\right)
}{
\sqrt{6}\,
(2\pi)^{\frac{1}{3-n}}
\left[
G_{\mathrm{eff}}M^{2}M_{\mathrm{Pl}}^{2}
\left(
1-e^{-\sqrt{\frac23}\frac{\phi}{M_{\mathrm{Pl}}}
}
\right)^{2}
\right]^{\frac{1}{3-n}}
}.
\end{equation}
The number of e-folds is obtained from Eq.~(\ref{N1}). For the  Starobinsky model, we find
\begin{equation}
N =
\frac{3(3-n)}{4}
(2\pi)^{\frac{2}{3-n}}
G_{\rm eff}^{\frac{2}{3-n}}
M^{\frac{2(n-1)}{3-n}}
M_{\rm Pl}^{\frac{4}{3-n}}
\left[
\mathcal{F}(\phi_c)-\mathcal{F}(\phi_f)
\right],
\end{equation}
where
\begin{equation}
\mathcal{F}(\phi)
=
\left(
1-e^{-\sqrt{\frac{2}{3}}\frac{\phi}{M_{\rm Pl}}}
\right)^{\frac{4}{3-n}}
{}_2F_1\left(
\frac{4}{3-n},
2;
\frac{7-n}{3-n};
1-e^{-\sqrt{\frac{2}{3}}\frac{\phi}{M_{\rm Pl}}}
\right),
\end{equation}
with ${}_2F_1$ being the Gauss hypergeometric function.

We now turn to the
slow-roll parameters.
Using the general
slow-roll relations derived in Sec.~\ref{SRI} together with the expressions
obtained above, the first slow-roll parameter takes the form
\begin{equation}
\varepsilon=\frac{
M^{2}\,
2^{\frac{2}{n-3}+1}\,
\pi^{\frac{2}{n-3}}\,
\exp\left(-2\sqrt{\frac{2}{3}}\frac{\phi}{M_{\mathrm{Pl}}}\right)
}{
3(3-n)
}  \left[
G_{\mathrm{eff}}M^{2}M_{\mathrm{Pl}}^{2}
\left(
\exp\left(-\sqrt{\frac{2}{3}}\frac{\phi}{M_{\mathrm{Pl}}}\right)-1
\right)^{2}
\right]^{\frac{2}{n-3}}.
\label{epsnew}
\end{equation}
The end of inflation is determined by $\varepsilon(\phi_f)=1$.
For generic $n$ this equation cannot be inverted in a simple closed
analytical form. We therefore consider small deviations from the
standard Starobinsky limit by setting
$n=1+\Delta$, with $|\Delta|\ll1$, and parameterize
$\gamma=\mu^{\,n-1}$, where $\mu$ is a fixed mass scale. To the leading order,
we then obtain
\begin{equation}
\label{phifnew}
\phi_f =
-\sqrt{\frac{3}{2}}\,M_{\rm Pl}
\ln(2\sqrt{3}-3)
+
\frac{2\sqrt{3}-3}{\sqrt{2}}M_{\rm Pl}
\left[
2+\ln\left(\frac{(2+\sqrt{3})\mu}{M}\right)
\right]\Delta\,.
\end{equation}
The reference scale $\mu$ is not independently fixed by the MHR
background dynamics and therefore reflects a residual
normalization freedom associated with the dimensional parameter
$\gamma$ away from the standard limit $n=1$. Since the Starobinsky
mass $M$ is the only intrinsic mass scale entering the
inflationary potential, it is natural to tie the MHR reference
scale to the inflationary sector by setting $\mu=c\,M$, with $c$ a
dimensionless constant of order unity. This choice avoids
introducing an additional, unrelated mass scale into the problem.
The remaining dimensionless normalization freedom can then be
fixed conveniently as $\mu=\frac{e^{-3/2}}{2+\sqrt{3}}\,M$. We
stress this prescription should be understood as a normalization
convention within the MHR parametrization, rather than as an
additional physical constraint.

As a result, the perturbative expression for the end-of-inflation
field takes the simple form
\begin{equation}
\phi_f =
-\sqrt{\frac{3}{2}}\,M_{\rm Pl}
\ln\left(2\sqrt{3}-3\right)
+
\frac{(2\sqrt{3}-3)}{2\sqrt{2}}\,
\Delta M_{\rm Pl}.
\label{eq:phi_f_simple}
\end{equation}
On the other hand, using Eq. \eqref{epet}, the second slow-roll parameter becomes
\begin{equation}
\eta=
\frac{M^2}{3}
e^{-\sqrt{\frac{2}{3}}\frac{\phi}{M_{\rm Pl}}}
\left(
2e^{-\sqrt{\frac{2}{3}}\frac{\phi}{M_{\rm Pl}}}-1
\right)
(2\pi)^{\frac{2}{n-3}}
\left[
G_{\rm eff}M^2M_{\rm Pl}^2
\left(
1-e^{-\sqrt{\frac{2}{3}}\frac{\phi}{M_{\rm Pl}}}
\right)^2
\right]^{\frac{2}{n-3}} .
\label{eq:eta_starobinsky_exact}
\end{equation}
\subsection{Inflationary dynamics and observational predictions}
We now determine the inflationary observables at horizon crossing.
Since the end-of-inflation field value has been obtained
perturbatively around the standard Starobinsky limit, we
consistently work to first order in $n=1+\Delta, |\Delta|\ll1$,
with $\gamma=\mu^\Delta$ and with the normalization of $\mu$ fixed
above.

For our purposes, it is convenient to introduce
\begin{equation}
x(\phi)\equiv
\exp\left(
-\sqrt{\frac{2}{3}}\frac{\phi}{M_{\rm Pl}}
\right).
\label{xdefinition}
\end{equation}
The expansion of Eq.~\eqref{epsnew} to first order in $\Delta$ gives
\begin{equation}
\epsilon(x)=
\frac{4}{3}\frac{x^2}{(1-x)^2}
\left\{
1+\Delta
\left[
\frac{1}{2}
-\ln\left(\frac{2+\sqrt{3}}{2}\right)
-\ln(1-x)
\right]
\right\}.
\label{epsDeltaexp}
\end{equation}
Similarly, expanding Eq.~\eqref{eq:eta_starobinsky_exact}, we obtain
\begin{equation}
\eta(x)=
\frac{4}{3}
\frac{x\left(2x-1\right)}{(1-x)^2}
\left\{
1+\Delta
\left[
-\ln\left(\frac{2+\sqrt{3}}{2}\right)
-\ln(1-x)
\right]
\right\}.
\label{etaDeltaexp}
\end{equation}
In order to evaluate the slow-roll parameters at horizon crossing,
we next determine the corresponding field value from the e-folds
relation. Using Eq. \eqref{N1} for the Starobinsky potential, we find
\begin{equation}
N=
\frac{9}{4}M_{\rm Pl}^2
\left(
\frac{8\pi G_{\rm eff}}{3}
\right)^{\frac{2}{3-n}}
\left(
\frac{3}{4}M^2M_{\rm Pl}^2
\right)^{\frac{2}{3-n}-1}
\int_{x_c}^{x_f}
\frac{(1-x)^{\frac{n+1}{3-n}}}{x^2}\,dx\,,
\label{Nstarx}
\end{equation}
which, upon expanding to leading order in $\Delta\ll1$, becomes
\begin{equation}
N=
\frac{3}{4}
\left\{
I_0+
\Delta
\left[
\ln\left(\frac{2+\sqrt{3}}{2}\right)I_0
+I_1
\right]
\right\}.
\label{NDelta}
\end{equation}
Here, we have defined
\begin{equation}
I_0=
\frac{1}{x_c}
-\frac{1}{x_f}
+\ln\left(\frac{x_c}{x_f}\right),\qquad
I_1=
\int_{x_c}^{x_f}
\frac{(1-x)\ln(1-x)}{x^2}\,dx\,,
\label{I1}
\end{equation}
where $x_{c\, (f)}\equiv x(\phi_{c\,(f)})$ and $\phi_f$ is consistently given by Eq. \eqref{eq:phi_f_simple}.

In analogy with the power-law analysis, we focus on the large-$N$
inflationary branch characterized by
$\phi_c\gg\phi_f$. From Eq. \eqref{xdefinition}, this hierarchy implies $x_c\ll x_f$.
Moreover, using Eq.~\eqref{eq:phi_f_simple}, the end-of-inflation value can be written as
\begin{equation}
x_f=
x_f^{(0)}
\left[
1-
\frac{x_f^{(0)}}{2\sqrt{3}}\,\Delta
\right]
+\mathcal{O}(\Delta^2)\,,
\label{xfDelta}
\end{equation}
where $x_f^{(0)}\equiv2\sqrt{3}-3$.
Therefore, $x_f\sim\mathcal{O}(1)$ in the perturbative regime.
Since $x_c\ll x_f$, the large-$N$ branch is consequently
characterized by $x_c\ll1$.

To determine the dominant contribution to the e-folds relation,
we first examine the behavior of the two integrals entering
Eq.~\eqref{NDelta}. The integral $I_1$ can be evaluated in closed
form as
\begin{equation}
I_1=
{\cal G}(x_f)-{\cal G}(x_c),\qquad
{\cal G}(x)=
-\ln x+\ln(1-x)
+\operatorname{Li}_2(x)
-\frac{\ln(1-x)}{x}.
\label{Gfunction}
\end{equation}
For $x_c\ll1$, the function ${\cal G}$ behaves as ${\cal G}(x_c)
=
1-\ln x_c
+\mathcal{O}(x_c)$,
and hence
$I_1=
\ln x_c
+{\cal G}(x_f)-1
+\mathcal{O}(x_c)$. We notice that, in the last equation, $x_f$ must be evaluated at zeroth order because $I_1$
already multiplies $\Delta$ in Eq.~\eqref{NDelta}.

The $x_f$-dependent part of $I_0$, on the other hand, must be
expanded explicitly:
\begin{equation}
-\frac{1}{x_f}-\ln x_f
=
-\frac{1}{x_f^{(0)}}-\ln x_f^{(0)}
+
\frac{x_f^{(0)}-1}{2\sqrt{3}}\,\Delta
+\mathcal{O}(\Delta^2).
\label{xfContribution}
\end{equation}
Substituting
into Eq.~\eqref{NDelta}, we obtain
\begin{equation}
\frac{4N}{3}
=
\frac{1}{x_c}
+\ln x_c
-\frac{1}{x_f^{(0)}}
-\ln x_f^{(0)}+
\Delta
\Bigg\{
\ln\left(\frac{2+\sqrt{3}}{2}\right)
\left[
\frac{1}{x_c}
+\ln x_c
-\frac{1}{x_f^{(0)}}
-\ln x_f^{(0)}
\right]
+\ln x_c
+{\cal G}\left(x_f^{(0)}\right)-1
+\frac{x_f^{(0)}-1}{2\sqrt{3}}
\Bigg\}.
\label{NrelationNLO}
\end{equation}
We now invert Eq.~\eqref{NrelationNLO} perturbatively in both
$\Delta$ and $1/N$. Retaining the terms required to determine
$x_c$ through next-to-leading order in the large-$N$ expansion, we
find
\begin{align}
x_c&=
\frac{3}{4N}
\left[
1+
\ln\left(\frac{2+\sqrt{3}}{2}\right)\Delta
\right]+
\frac{9}{16N^2}
\Bigg\{
-\ln\left(\frac{4N}{3}\right)
-\frac{1}{x_f^{(0)}}
-\ln x_f^{(0)}
+\Delta
\Bigg[
-\left(
1+
2\ln\left(\frac{2+\sqrt{3}}{2}\right)
\right)
\ln\left(\frac{4N}{3}\right)
\nonumber\\[2mm]
&\qquad
-2\ln\left(\frac{2+\sqrt{3}}{2}\right)
\left(
\frac{1}{x_f^{(0)}}+\ln x_f^{(0)}
\right)
+\ln\left(\frac{2+\sqrt{3}}{2}\right)
+\frac{x_f^{(0)}-1}{2\sqrt{3}}
+{\cal G}\left(x_f^{(0)}\right)-1
\Bigg]
\Bigg\}
+
\mathcal{O}\left(
\frac{\ln^2N}{N^3},
\frac{\Delta\ln^2N}{N^3},
\frac{\Delta^2}{N}
\right).
\label{xcNLO}
\end{align}
Since $\epsilon_c\sim\mathcal{O}(N^{-2})$, the
next-to-leading contribution in Eq.~\eqref{xcNLO} affects
$\epsilon_c$ only beyond the order required here. We therefore
obtain
\begin{equation}
\epsilon_c=
\frac{3}{4N^2}
\left\{
1+
\left[
\frac{1}{2}
+
\ln\left(\frac{2+\sqrt{3}}{2}\right)
\right]\Delta
\right\}\,.
\label{epsilonCrossNLO}
\end{equation}
For $\eta_c$, instead, the next-to-leading correction to $x_c$
must be retained because $\eta_c=\mathcal{O}(N^{-1})$. Expanding
Eq.~\eqref{etaDeltaexp}, we obtain
\begin{align}
\eta_c
=&
-\frac{1}{N}
+
\frac{3}{4N^2}
\left[
\ln\left(\frac{4N}{3}\right)
+\frac{1}{x_f^{(0)}}
+\ln x_f^{(0)}
\right]+
\frac{3\Delta}{4N^2}
\Bigg\{
\left[
1+
\ln\left(\frac{2+\sqrt{3}}{2}\right)
\right]
\ln\left(\frac{4N}{3}\right)
\nonumber\\
&
+\ln\left(\frac{2+\sqrt{3}}{2}\right)
\left[
\frac{1}{x_f^{(0)}}+\ln x_f^{(0)}
\right]
-\ln\left(\frac{2+\sqrt{3}}{2}\right)
-\frac{x_f^{(0)}-1}{2\sqrt{3}}
-{\cal G}\left(x_f^{(0)}\right)
\Bigg\}.
\label{etaCrossNLO}
\end{align}
We observe that the terms proportional to $\Delta/N$ cancel exactly. Thus, the
first nonvanishing linear MHR correction to $\eta_c$ appears at
order $\Delta/N^2$.

Following the phenomenological prescription adopted in
Sec.~\ref{SRI}, we evaluate the inflationary observables using
Eqs. \eqref{rns1} and \eqref{ns}. The tensor-to-scalar ratio
therefore becomes
\begin{equation}
r=
\frac{12}{N^2}
\left\{
1+
\left[
\frac{1}{2}
+
\ln\left(\frac{2+\sqrt{3}}{2}\right)
\right](n-1)
\right\},
\label{rStarNLO}
\end{equation}
while, for the scalar spectral index, we obtain
\begin{align}
n_s
={}&
1-\frac{2}{N}
+
\frac{1}{N^2}
\Bigg\{
\frac{3}{2}
\left[
\ln\left(\frac{4N}{3}\right)
+\frac{1}{x_f^{(0)}}
+\ln x_f^{(0)}
\right]
-\frac{9}{2}
\Bigg\}+
\frac{n-1}{N^2}
\Bigg\{
\frac{3}{2}
\Bigg[
\left(
1+
\ln\left(\frac{2+\sqrt{3}}{2}\right)
\right)
\ln\left(\frac{4N}{3}\right)
\nonumber\\
&\qquad
+\ln\left(\frac{2+\sqrt{3}}{2}\right)
\left(
\frac{1}{x_f^{(0)}}+\ln x_f^{(0)}
\right)
-\ln\left(\frac{2+\sqrt{3}}{2}\right)
-\frac{x_f^{(0)}-1}{2\sqrt{3}}
-{\cal G}\left(x_f^{(0)}\right)
\Bigg]
-\frac{9}{2}
\left[
\frac{1}{2}
+
\ln\left(\frac{2+\sqrt{3}}{2}\right)
\right]
\Bigg\}.
\label{nsStarNLO}
\end{align}
As a consistency check, one can verify that the standard Starobinsky expressions are recovered for $n\to1$.

\begin{figure}[t]
\centering
    \includegraphics[width=0.5\linewidth]{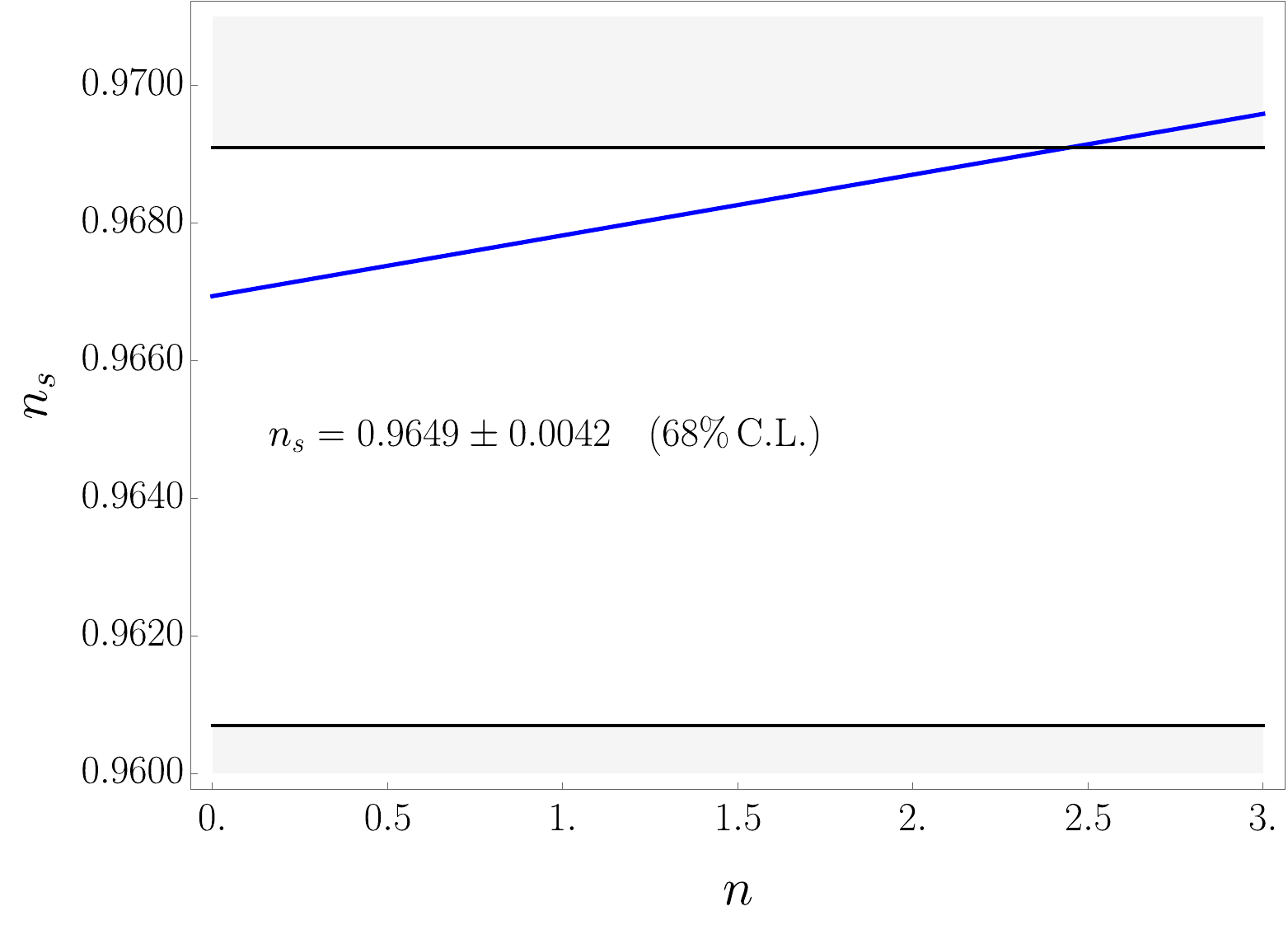}
\caption{Scalar spectral index $n_s$ (blue line) as a function of the MHR parameter $n$
for $N=60$. The shaded regions denote the parameter space excluded
by the observational constraints (black solid lines). The curve is obtained from the
first-order expansion around $n=1$ and, away from the perturbative
regime $|n-1|\ll1$, should be understood as a formal extrapolation
of the perturbative prediction.}
\label{Fig2}
\end{figure}

Let us now compare the above predictions with the observational bounds
adopted in Sec.~\ref{SRI}. For $N=60$, imposing $n>0$, the Planck
determination $n_s=0.9649\pm0.0042$ at $68\%$ C.L. formally yields
$n\lesssim2.449$ (see also Fig.~\ref{Fig2}). The tensor bound $r<0.036$,
on the other hand, formally gives the considerably weaker constraint
$n\lesssim9.720$ and therefore does not further restrict the parameter
space allowed by the scalar spectral index. Thus, in contrast with the
power-law case, the Starobinsky potential exhibits a nontrivial
observational sensitivity to the MHR parameter, leading to a formal
upper constraint on $n$. However, since the present analysis is perturbative
around $n=1$, with $|n-1|\ll1$, the numerical value of this upper limit
extends beyond the strictly controlled regime of the expansion and
should therefore be understood as an indication of the constraining
power of $n_s$, rather than as a global constraint on the full MHR
parameter space.

\subsection{Power Spectrum}
The perturbative analysis can provide additional
observational information through the normalization of the scalar
power spectrum. In particular, the amplitude of scalar
perturbations, $A_s$, determines how the normalization scale of
the Starobinsky potential is modified by the MHR deformation.
We therefore turn to the study of $A_s$.

Consistently with the
phenomenological treatment of $r$ and $n_s$, we adopt the standard
canonical expression \cite{Baumann}
\begin{equation}
A_s\equiv{\cal P}_{\cal R}(k_c)
\simeq
\frac{H_c^2}{8\pi^2M_{\rm Pl}^2\epsilon_c}.
\label{AsPrescription}
\end{equation}
Expanding the modified Hubble rate \eqref{Fr3bis} to first order in
$\Delta$, we find
\begin{equation}
H^2(x)=
\frac{M^2}{4}(1-x)^2
\left\{
1+\Delta
\left[
\ln\left(\frac{2+\sqrt{3}}{2}\right)
+\ln(1-x)
\right]
\right\},
\label{H2Delta}
\end{equation}
where $x$ has been defined in Eq. \eqref{xdefinition}. Combining
Eqs.~\eqref{H2Delta} and \eqref{epsDeltaexp}, we obtain
\begin{align}
A_s(x)
=
\frac{3M^2}{128\pi^2M_{\rm Pl}^2}
\frac{(1-x)^4}{x^2}
\left\{
1+\Delta
\left[
2\ln\left(\frac{2+\sqrt{3}}{2}\right)
+2\ln(1-x)
-\frac{1}{2}
\right]
\right\},
\label{Asx}
\end{align}
which can be evaluated at horizon crossing
to give
\begin{equation}
A_s(N)
=
\frac{M^2N^2}{24\pi^2M_{\rm Pl}^2}
\Bigg\{
1-\frac{\Delta}{2}+
\frac{1}{N}
\Bigg[
\frac{3}{2}
\left(
\ln\left(\frac{4N}{3}\right)
+\frac{1}{x_f^{(0)}}
+\ln x_f^{(0)}
\right)
-3
+\Delta\,{\cal A}_1(N)
\Bigg]
\Bigg\}\,,
\label{AsNLO}
\end{equation}
where
\begin{align}
{\cal A}_1(N)
=\,&
\frac{3}{2}
\ln\left(\frac{2+\sqrt{3}}{2}\right)
\left[
\ln\left(\frac{4N}{3}\right)
+\frac{1}{x_f^{(0)}}
+\ln x_f^{(0)}
\right]-
\frac{9}{2}
\ln\left(\frac{2+\sqrt{3}}{2}\right)
-\frac{3}{4}
\left[
\frac{1}{x_f^{(0)}}
+\ln x_f^{(0)}
\right]
\nonumber\\
&+
\frac{3}{4}\ln\left(\frac{4N}{3}\right)
-\frac{3}{2}{\cal G}\left(x_f^{(0)}\right)
-
\frac{\sqrt{3}}{4}x_f^{(0)}
+\frac{\sqrt{3}}{4}
+\frac{3}{2}\,.
\label{A1definition}
\end{align}
It is straightforward to verify that, at leading order in the
large-$N$ expansion and for $n=1$ (i.e., $\Delta=0$), Eq.
\eqref{AsNLO} reduces to the standard Starobinsky prediction
$A_s^{\rm St} \simeq \frac{M^2N^2}{24\pi^2M_{\rm Pl}^2}$ (see,
e.g., Ref. \cite{Ivanov:2021chn}). In the standard Starobinsky
model, the mass parameter $M$ fixes the overall normalization of
the inflationary potential and is determined by matching the
scalar amplitude to its observed value. In the present framework,
however, $A_s$ depends on both $M$ and $n$, so that, if $M$ is
allowed to vary, the observed amplitude determines a relation
$M=M(n)$ rather than an independent constraint on $n$.

To make this statement explicit, it is convenient to rewrite
Eq.~\eqref{AsNLO} as
\begin{equation}
A_s(N)=
\frac{M^2N^2}{24\pi^2M_{\rm Pl}^2}
\left[
{\cal K}_0(N)
+(n-1){\cal K}_1(N)
\right],
\label{AsCompact}
\end{equation}
where
\begin{eqnarray}
\label{F0definition}
{\cal K}_0(N)
&=&
1+
\frac{1}{N}
\left[
\frac{3}{2}
\left(
\ln\left(\frac{4N}{3}\right)
+\frac{1}{x_f^{(0)}}
+\ln x_f^{(0)}
\right)
-3
\right],\\[2mm]
{\cal K}_1(N)
&=&
-\frac{1}{2}
+\frac{{\cal A}_1(N)}{N}.
\label{F1definition}
\end{eqnarray}
Here we have used $\Delta=n-1$.

For fixed $N$, imposing the observed scalar amplitude gives
\begin{equation}
\frac{M(N,n)}{M_{\rm Pl}}
=
\frac{1}{N}
\left[
\frac{24\pi^2 A_s^{\rm obs}}
{{\cal K}_0(N)+(n-1){\cal K}_1(N)}
\right]^{1/2}\,,
\label{Mofn}
\end{equation}
which, in the general case, provides a  non-trivial relation
between the Starobinsky normalization scale and the MHR
deformation parameter.

However, it is instructive to consider a complementary
fixed-normalization scenario. In other terms, one may regard the
MHR modification as a deformation of the cosmological background
dynamics while keeping the underlying Starobinsky inflationary
sector fixed. Such a prescription allows us to isolate the effect
of the MHR deformation from a simultaneous retuning of the
inflationary potential normalization.

Within this framework, let $M_{\rm St}(N)$ denote the value of the
Starobinsky mass parameter that reproduces the central observed
scalar amplitude for $n=1$, evaluated consistently at the same
order in the large-$N$ expansion as Eq.~\eqref{AsNLO}. We then
have
\begin{equation}
A_{s,0}^{\rm obs}
=
\frac{M_{\rm St}^2N^2}{24\pi^2M_{\rm Pl}^2}
{\cal K}_0(N),
\label{MStdefinition}
\end{equation}
where $A_{s,0}^{\rm obs}$ denotes the central observational value.

If the Starobinsky normalization is kept fixed under the MHR
deformation, $M=M_{\rm St}$,
Eq.~\eqref{AsCompact} gives
\begin{equation}
\frac{A_s(N,n)}{A_{s,0}^{\rm obs}}
=
1+
\frac{{\cal K}_1(N)}{{\cal K}_0(N)}
(n-1).
\label{AsRatio}
\end{equation}
Thus, once the normalization scale is held fixed, the departure
of the scalar amplitude from its Starobinsky value is controlled
directly by the MHR parameter.

Now, for the representative value $N=60$, we obtain
\begin{equation}
\frac{A_s(60,n)}{A_{s,0}^{\rm obs}}
\simeq
1-0.403\,(n-1).
\label{AsRatio60}
\end{equation}
Since the whole analysis is performed to first order in
$\Delta$, taking the logarithm consistently gives
\begin{equation}
\ln\left[
\frac{A_s(60,n)}{A_{s,0}^{\rm obs}}
\right]
\simeq
-0.403\,(n-1)
+\mathcal{O}\left((n-1)^2\right).
\label{lnAsRatio60}
\end{equation}
Within the phenomenological prescription adopted above, the
observational uncertainty on the scalar amplitude can therefore be
used to estimate the sensitivity of the fixed-normalization
scenario to departures from the Starobinsky limit. Using the
Planck value \cite{Planck2018VI}
\begin{equation}
\ln(10^{10}A_s)=3.045\pm0.016
\qquad (68\%~{\rm C.L.}),
\label{PlanckAs}
\end{equation}
and recalling that $M_{\rm St}$ is fixed by requiring the $n=1$
prediction to reproduce the central observed value, we have
$\ln(10^{10}A_{s,0}^{\rm obs})=3.045$. It follows that the
observational interval can equivalently be written as
\begin{equation}
\left|
\ln(10^{10}A_s)
-
\ln(10^{10}A_{s,0}^{\rm obs})
\right|
\lesssim0.016.
\end{equation}
Using Eq. \eqref{lnAsRatio60}, we finally obtain
\begin{equation}
\label{inter}
0.403\,|n-1|
\lesssim0.016\,\,\, \Longrightarrow\,\,\, 0.960\lesssim n\lesssim1.040\,.
\end{equation}
This result shows that, within the fixed-normalization scenario,
the scalar amplitude is significantly more sensitive to the MHR
deformation than the spectral observables considered above. We
stress, however, that the interval \eqref{inter} should be
interpreted as a conditional phenomenological constraint, as it
relies on the assumption that the Starobinsky normalization scale
is kept fixed, $M=M_{\rm St}$, so that the change in $A_s$ is
attributed entirely to the MHR deformation.

Before concluding, it is interesting to place our result in the
broader context of existing observational constraints for the
generalized MHR entropy. In this regard, we observe that our
result \eqref{inter} is competitive with several bounds previously
obtained from non-inflationary observables. In particular, the
baryogenesis analysis presented in \cite{Luciano:2025fqg} gives
$0.98\lesssim n<1$, while the combination SNIa+CC+BAO(DESI
DR2)+SH0ES yields $n=0.945\pm0.070$ \cite{Luciano:2025ovj}.
Furthermore, primordial gravitational waves provide the lower
bound  $n\gtrsim0.884$ \cite{Luci1}, whereas the SNIa+CC+BAO(DESI
DR1) analysis gives $n=1.09\pm0.01$ \cite{Basilakos:2025wwu}.

Our inflationary constraint is therefore more restrictive than the
DESI DR2+SH0ES interval and the primordial-gravitational-wave bound,
while being of comparable order to the baryogenesis constraint.
The DESI DR1 determination instead favors values above unity and
does not overlap with the interval obtained here at the quoted
uncertainty. These comparisons should nevertheless be interpreted
with some caution, since the different bounds probe distinct
cosmological epochs and rely on different assumptions concerning
the remaining parameters of the model.

For completeness, we mention that broader classes of generalized horizon entropies
have also recently been confronted with cosmological observations
\cite{Nojiri:2023bom,Prasanthan:2026boc,Prasanthan:2026vyk,Leizerovich:2026pfy}. Since these frameworks involve more general entropy
functionals and additional parameter dependencies, their constraints
are not in one-to-one correspondence with the bound on $n$ obtained
here and a direct numerical comparison would be model dependent. Nevertheless,
they  provide a complementary demonstration of the
ability of cosmological observations to probe departures from
standard horizon thermodynamics and, in a broader sense, from standard gravitational theory.
In this respect, our result offers
an independent probe from the inflationary epoch, directly
constraining the deformation of the mass--horizon relation considered
in the present work.
\section{Closing remarks \label{Con}}
In this work, we have investigated slow-roll inflation in a
modified cosmological framework grounded in a generalized MHR.
Using Padmanabhan's emergence paradigm,  we have obtained the
modified Friedmann dynamics using generalized MHR and explored its
implications for inflation driven by a canonical scalar field.

Our analysis of power-law potentials revealed that the MHR-induced
modification of the background expansion, while altering the
slow-roll parameters and inflationary observables, does not yield
a common parameter region compatible with the combined CMB
constraints on $n_s$ and $r$ for the representative monomial
models considered. Thus, within the present MHR realization, the
observational tension plaguing these models in the standard
scenario persists, indicating that the deformation of horizon
thermodynamics considered here is insufficient to rescue power-law
inflation.

A markedly different picture emerged for the Starobinsky model.
Working perturbatively around the Bekenstein-Hawking limit $n = 1
+ \Delta$, we derived the modified inflationary dynamics and the
corresponding observables in the large-$N$ regime. A key result of
our analysis concerns the scalar power-spectrum normalization
$A_s$. When the MHR modification is interpreted as a deformation
of the cosmological background while keeping the underlying
Starobinsky normalization scale fixed, $A_s$ provides a remarkably
sensitive probe of the MHR exponent. For $N = 60$, this
prescription yielded the constraint $0.960 \lesssim n \lesssim
1.040$, which is significantly tighter than the bounds derived
from the spectral observables $r$ and $n_s$, and is competitive
with existing constraints from baryogenesis and large-scale
structure probes. This highlights that the amplitude of primordial
perturbations is particularly susceptible to departures from the
standard $n = 1$ scaling, offering a powerful observational handle
on generalized horizon thermodynamics.

Beyond the specific constraints obtained, our results reveal a
nontrivial and intricate interplay between horizon entropy
modifications and primordial inflation. They demonstrate that
inflationary observables can serve as sensitive discriminators of
modifications to the MHR scaling, complementing probes from
late-time cosmology, structure formation, and gravitational-wave
physics. The sensitivity of $A_s$ to the MHR deformation
underscores the importance of precision measurements of the
primordial power spectrum in testing fundamental aspects of
gravitational thermodynamics.

Several important directions remain open for future investigation.
A natural and immediate next step is the development of a complete
perturbation theory for MHR cosmology. Deriving the scalar and
tensor mode equations directly within the modified framework would
remove the need for the phenomenological perturbation prescription
adopted in this work, allowing the primordial spectra, their
normalization, and their scale dependence to be determined
self-consistently. This would provide a crucial test of whether
the sensitivity found at the background level persists at the
level of cosmological perturbations, further strengthening the
potential of inflation as a probe of generalized horizon
thermodynamics. Additionally, extending the analysis to more
general entropy functionals and exploring the interplay between
the MHR deformation and other quantum-gravity-inspired corrections
would help place our findings in a broader theoretical context.
Furthermore, it would be interesting to investigate whether the
MHR modification affects the reheating epoch and subsequent
cosmological evolution, potentially leaving observable imprints in
the cosmic microwave background and large-scale structure that
could complement or refine the constraints obtained here. A
systematic investigation along these directions is currently
underway and will be presented in a future work.

Overall, our study establishes that inflation, and particularly
the Starobinsky model, offers a sensitive and complementary window
into departures from standard horizon thermodynamics, with the
scalar power-spectrum normalization emerging as a particularly
powerful observable for constraining modifications to the MHR
scaling.

\acknowledgments{The work of A. Sheykhi is based upon research
funded by Shiraz University under project No. 5RLG2M82894. The
research of GGL is supported by the postdoctoral fellowship
program of the University of Lleida. GGL gratefully acknowledges
the contribution of the LISA Cosmology Working Group (CosWG), as
well as support from the COST Actions CA21136 - \textit{Addressing
observational tensions in cosmology with systematics and
fundamental physics (CosmoVerse)} - CA23130, \textit{Bridging high
and low energies in search of quantum gravity (BridgeQG)} and
CA21106 - \textit{COSMIC WISPers in the Dark Universe: Theory,
astrophysics and experiments (CosmicWISPers)}.}

\end{document}